# Non-reciprocal second harmonic generation in ferroelectric photonic structures

Pablo Molina, Mariola O Ramírez, Basilio J. García* and Luisa E. Bausá

Departamento de Física de Materiales and *Departamento de Física Aplicada
Universidad Autónoma de Madrid
28049-Madrid, Spain.

**Second harmonic generation is a powerful tool directly connected to the symmetry of materials. Phase transitions, lattice rotations or electromagnetic coupling in multiferroic compounds can be revealed by using second harmonic generation.[1-4] Here we show that the sense of the spontaneous electric polarization $P_s$ in trigonal ferroelectrics can lead to a pronounced spatial dependence of the second harmonic generation. By using a two-dimensional nonlinear photonic structure we demonstrate that counter-propagating beams along the polar axis generate non-reciprocal second harmonic patterns. In optics, with some exception,[5] non-reciprocal phenomena are generally related to the breaking of time-reversal symmetry involving a magnetic field.[6,7] Here, in contrast, the non-reciprocity is inherent to the quadratic nonlinear tensor being related to the inversion symmetry breaking at the ferroelectric phase. These findings provide novel routes to generate nonreciprocal light-matter interaction processes in two-dimensional structures assembled onto polar surfaces,[8,9] including metals for plasmonics or biological compounds.**

To reveal non-reciprocal second harmonic generation (SHG) processes, non-symmetric SHG spatial patterns are required. The simplest case corresponds to patterns lacking axial symmetry with respect to a fundamental beam propagating along the polar axis. For this aim we have fabricated a two dimensional nonlinear photonic crystal (2D-NLPC) in $LiNbO_3$ consisting of a periodic arrangement of ferroelectric domains with alternating orientation of $P_s$ (See methods). The distribution of domains produces a modulation of the sign of the second order nonlinear susceptibility $\chi^{(2)}$ providing a set of reciprocal lattice vectors of the structure of domains in the plane of modulation, $xy$.[10] As a result, SHG occurs at multiple directions and nonlinear Bragg diffraction patterns can be obtained.[11-14] We have chosen a NLPC in which the total inverted area only comprises a 3 % of the overall sample area ($f\sim 0.03$), so that the original orientation of $\mathbf{P}_s$ is largely dominant (Fig. 1a).

Figure 1(b and c) show a scheme of the SHG processes when a fundamental beam propagates parallel to the ferroelectric axis, together with the far field SHG pattern experimentally obtained. In this configuration all the reciprocal vectors lie on the plane perpendicular to the fundamental wavevector $\mathbf{k}_\omega$. Therefore, for a fixed frequency, $\omega$, the SHG waves propagate with a conical angle $\theta$ between the wavevectors $\mathbf{k}_\omega$ and $\mathbf{k}_{2\omega}$ defined by the momentum conservation law, which forces the particular angular dispersion $\cos\theta = 2k_\omega/k_{2\omega} = n^o(\omega)/n^{o,e}(2\omega)$ where $n^{o,e}(\omega)$ is the ordinary/extraordinary refractive index of the material at a frequency $\omega$. Accordingly, our pattern corresponds to the far field of the beams generated inside the crystal by the participation of multiple order nonlinear Bragg diffraction, which are all distributed onto a circular ring for a particular color. The ring shaped SHG pattern shows a hexagonal distribution of more intense and better defined and regions which are related to the nonlinear diffraction produced by the hexagonal symmetry of the ferroelectric domain walls (Fig. 1c).

The generated pattern exhibits two main features: a marked azimuthal dependence of the SHG intensity, and the co-existence of two SHG rings. Both characteristics are related to the anisotropy of $LiNbO_3$. In this birefrigent material a beam propagating at an angle $\theta$ with respect to the optical axis can be decomposed in two different components: one traveling along the fast axis ($n^e$) and a second one traveling along the slow ($n^o$) axis. Then, two types of SHG processes [$e,oo$] and [$o,oo$] (two ordinary waves generate an extraordinary/ordinary SH wave) take place simultaneously and can be separated because of the difference in their angular dispersion, $\Delta\theta$, which is directly proportional to $\Delta n(2\omega)$, (See methods). Additionally, the patterns related to each process show a quite different azimuthal intensity profile. For the $3m$ point group symmetry of $LiNbO_3$ their azimuthal intensity profiles can be written in terms of the nonlinear effective coefficients $d_{eff}^{ord}$ and $d_{eff}^{ext}$ as:

$$I_{SHG}^{ord} \approx (d_{eff}^{ord})^2 = (d_{22}\cos(\varphi+2\gamma))^2$$
$$I_{SHG}^{ext} \approx (d_{eff}^{ext})^2 = (d_{31}\sin\theta + d_{22}\sin(\varphi+2\gamma)\cos\theta)^2$$
(1)

where $d_{22} = d_{yxx} = -d_{yyy}$ and $d_{31}=d_{zxx}$ are the nonlinear coefficients, $\varphi$ the azimuthal angle measured counterclockwise from the $x$ axis, and $\gamma$ the polarization angle of the linearly polarized fundamental incident beam measured from the $x$ axis ($\gamma = 0$ for polarization along the $x$ axis). The results of the simulation for each process, as well as the composition of both rings, are shown in Fig. 1(d). The far field pattern obtained for the ordinary process shows a twofold symmetry axis, while the extraordinary one lacks of axial symmetry. As observed, the intensity of the experimental SH nonlinear Bragg pattern is modulated by the azimuthal dependence of $d_{eff}$. In LiNbO$_3$ the value of $d_{31}$ is approximately twice larger than that of $d_{22}$.[15,16]

The non-reciprocal optical response has been revealed by analyzing the effect that the orientation of $P_s$ (↑ and ↓) produces on the SHG patterns. We have considered two different configurations. In the first one, we compare the nonlinear Bragg diffraction patterns produced by two different linearly polarized fundamental beams ($\gamma = 0$) propagating with opposite directions along the $z$ axis when they are launched onto the crystal with the $x$ axis directed along the vertical direction (Fig. 2). Upon this configuration the SHG far field images are identical, regardless the direction $+z$ and $-z$ of the fundamental incident (Fig. 2). The result is well reproduced by eq. (1) taking into account the changes in the sign of the nonlinear coefficients $d_{22}$ and $d_{31}$ when changing the crystal-fixed axis relative to the laboratory-fixed axis. It is consistent with a $C_2(x)$ rotation, which is equivalent to the effect of the electrical poling.[17] In LiNbO$_3$ reversing $P_s$ by applying an electric field along the $z$ axis produces not only a reversion of the cationic chain along the polar axis, but also a reversion in the chain of cations along the $y$ axis.[18] Accordingly, the polarization inversion can be seen as a $180^0$ rotation of the crystal structure around the $x$ crystallographic axis, $C_2(x)$, which leaves the actual crystal structure unchanged. The reciprocity observed in Fig. 2 shows that is not possible to distinguish the SHG patterns produced by two different counter-propagating fundamental beams when the nonlinear interaction experiences the physical changes on both $z$ and $y$ axis after the $C_2(x)$ rotation, which geometrically connect the two real polarization states $P_s$↑ and $P_s$↓ in the ferroelectric crystal.

However, it is possible to use a configuration in which the SHG pattern is sensitive to the direction of the propagation of the fundamental beam along the $z$ axis. Figure 3 shows the two far field SHG images obtained for two linearly polarized beams (again $\gamma=0$) propagating with opposite directions along the $z$ axis, impinging on the crystal with the $y$ axis on the vertical direction. As seen, the nonlinear Bragg diffraction differs intriguingly for light propagating in opposite directions: the orientation of the pattern changes from "up" to "down" when the fundamental beam propagation is changed from -$z$ to +$z$. The result is reproduced by eq. (1) taking into account the twofold rotation around the $y$ axis, $C_2(y)$, for which $d_{31}$ changes its sign, while $d_{22}$ remains unchanged. This operation changes the SHG pattern from "up" to "down" leading to a disparate nonlinear optical response for fundamental beams traveling with opposite directions along $z$. The observed non-reciprocity in the SHG process manifests the disparity of the two dominant orientations $\mathbf{P}_s \uparrow$ and $\mathbf{P}_s \downarrow$ when the light propagation direction is reversed along the polar axis. It is worth to mention that the 180º rotation around the $y$ axis does not lead to the actual physical domain inversion under electrical poling along the z axis, since it only changes the sign of the $z$ axis, leaving unchanged the sign of $y$. Consequently, the results of Fig. 3 show the different nonlinear interaction when only the orientation of the cations along the polar $z$ axis is reversed, since the $C_2(y)$ operation leaves intact the order of cations along the $y$ axis. The results are analog to what would happen in the crystal if solely the chain of cations along the $z$ axis would be changed after inverting the polarization. In fact, that would be equivalent to analyze the effect two ideal orientational states $\mathbf{P}_s \uparrow$ and $\mathbf{P}_s \downarrow$ in LiNbO$_3$ in which each state mirrors the other one through the plane perpendicular to the polar axis ($xy$), since the sense of $x$ and $y$ are not physically relevant after the $C_2(y)$ operation. Therefore, Fig. 3 simulates the results that could be obtained with two orientational states showing different "handedness" when probing through nonlinear Bragg diffraction.

The effect of the polarization of the fundamental beam on the reciprocity of the nonlinear Bragg patterns is shown in Fig.4. As derived from eq. (1), the azimuthal intensity profile of the generated SH waves depends on twice the polarization angle of input light, $2\gamma$. Hence, for fundamental beams linearly polarized along $x$ and $y$ directions, ($\gamma = 0, \pi/2$, respectively) it is possible to generate "complementary" SHG patterns (Fig. 4a). Consequently, the interaction of a circularly polarized fundamental beam with the crystal leads to reciprocal nonlinear diffraction patterns with two perpendicular symmetry planes (Fig. 4b). Therefore, it is possible to switch from a non-reciprocal to a reciprocal optical configuration by changing the polarization state of the fundamental beam into the crystal.

In conclusion, by using optical probes we demonstrate the possibility of "visualizing" the sense of the dominant $\mathbf{P}_s$ in a ferroelectric crystal to the naked eye. For the first time, we have revealed optical non-reciprocity on the distribution intensity of frequency conversion processes for fundamental counter propagating beams along the polar axis in a ferroelectric crystal. The work has been focused on LiNbO$_3$, but the results are applicable to other type ferroelectric systems. Additionally, since the non-reciprocal effects are extensible to a large variety of quadratic three-wave mixing processes, the observations can expand the multifunctional character of 2D-NLPC,[19-22] particularly when placed into optical cavities. Further, the non reciprocity obtained in the nonlinear optical response shows the potential of ferroelectric patterning as a valuable tool to generate novel spatially selective and non-reciprocal light-matter interaction processes in two-dimensional structures.

# FIGURE CAPTIONS

**FIGURE 1.** (a) Optical micrograph showing the 2D distribution of ferroelectric inverted domains on the +z cut of a $LiNbO_3$ crystal. Lines on the picture are a guide to show the hexagonal symmetry of the domain pattern. The domain walls also show hexagonal shape according to the crystal symmetry of the $LiNbO_3$. (b) Schematic of the conical SHG process for an incident fundamental beam along the z axis of the crystal; The orientation of the dominant $P_s$ has been marked with an arrow. (c) Far field SH pattern generated at 600 nm when a fundamental beam at 1200 nm propagates parallel to the ferroelectric axis ($k_\omega \parallel z$). (d) Projection of the intensities of the ordinary (left), extraordinary (center) and combination of both (right) SGH beams on the XY plane of the screen, as follows from equation 1.

**FIGURE 2**. Far field nonlinear diffraction patterns observed when linearly polarized ($\gamma = 0$) counterpropagating beams are launched on the crystal with the *x* axis in vertical position. The result is consistent with the effect of the electrical poling along the *z* axis (analog to a $C_2(x)$ operation) which lead the patterns unchanged. The orientation of the dominant $P_s$ has been marked with an arrow.

**FIGURE 3**. Different SHG far field patterns obtained for two linearly polarized beams ($\gamma = 0$) propagating with opposite directions along the polar axis of the nonlinear photonic crystal with the *y* axis in vertical position. The orientation of the pattern is reversed from "up" to "down" when the fundamental propagation direction is changed from -z to +z. The result shows the non-reciprocity of the SHG process associated with the sense of the dominant spontaneous polarization $P_s$.

**FIGURE 4**. Effect of the polarization states of the fundamental beam on the SHG patterns. (a) Far field patterns obtained for two linearly polarized fundamental beams with polarization states parallel to *x* and *y* axis of the crystal. The distribution intensity of SHG, depicted at the right side of each pattern, has been calculated according to eq. 1 taking into account the changes in the polarization angle $\gamma$. (b) Far field pattern obtained for a circularly polarized fundamental beam.

# ACKNOWLEDGEMENTS

The authors gratefully acknowledge the support by the Spanish Ministry of Science and Innovation under contract MAT2007-64686, and the University Autonoma of Madrid under contract CCG08-UAM/MAT-4434

# Methods

Fabrication of the two dimensional nonlinear photonic structure

A 1 mm thick plate from a congruent ([Li]/[Nb]=0.945) single domain LiNbO$_3$ crystal doped with MgO (5%) was cut and polished with its main faces oriented perpendicular to the ferroelectric *z*-axis. The incorporation of MgO is useful for optical applications since it reduces the photorefractive effect in LiNbO$_3$. Additionally, the coercive field required for ferroelectric domain inversion is reduced in MgO doped LiNbO$_3$ crystals. To produce the NLPC we have employed direct electron beam writing (DEBW) on the -*z* face of the crystal by means of a Philips XL30 Schottky field emission gun electron microscope driven by an Elphy Raith nanolithography software. The irradiation process was performed without any mask. Before the electron bombardment, a 100 nm thin film of Al was evaporated onto the +*z* face of the sample, which acted as a ground electrode during the electron beam bombardment. The irradiation parameters were 10 keV of incident electron energy, 400 pA irradiation current and 1000 µC/cm$^2$ of electronic dose. The obtained anti-parallel ferroelectric domain structure was revealed after selective chemical etching in a 2:1 solution of HNO$_3$:HF at room temperature for 20 min.

The fabricated NLPC consisted of a two dimensional hexagonal array of hexagonal column-shaped polarization inverted domains ($P_s\downarrow$) embedded into a single domain LiNbO$_3$ crystal of opposite polarization ($P_s\uparrow$). The inverted domain columns were directed along the polar axis of the crystal (*z* axis) traversing the whole sample thickness (1 mm). The diameter of the inverted domain columns in the *xy* plane was 3 µm and the lattice parameter of the hexagonal array was Λ= 20 µm. The average filling factor, *f,* defined as the ratio of the area of the inverted region to the entire area, was ~ 0.03 in both, +*z* and -*z* faces. The spatial extension of the patterns use in this work was 2x2 mm$^2$.

Optical measurements

For the second harmonic generation (SHG) experiments, the sample was polished up to optical quality and mounted over a stage allowing the different orientations of the crystal with respect to the fundamental incident wave. The fundamental infrared radiation was provided by an optical parametric oscillator (Spectra Physics MOPO 730). This source generates pulses of 10 ns at a repetition rate of 10 Hz. The energy per pulse used in this work was around 10 mJ (1 MW peak power). Among the multi-wavelength processes allowed by the multiple orders, the red one was the most clearly visible to the naked eye since it corresponds to the lowest order in the visible region. For the fundamental wavelength used in this work ($\lambda_\omega$= 1200 nm), the external conical angles for the extraordinary and ordinary waves are $\theta_{ext}$=16.4° and $\theta_{ord}$=17.6°.

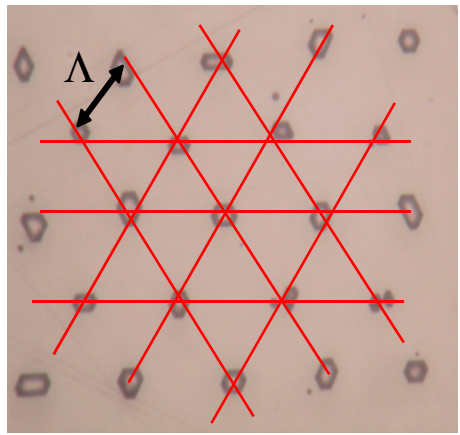
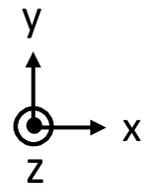
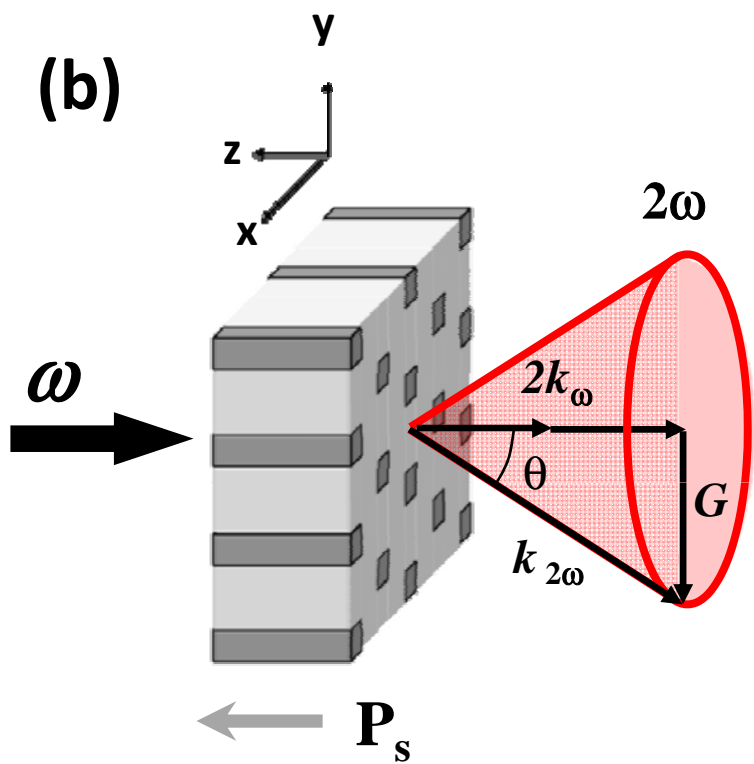
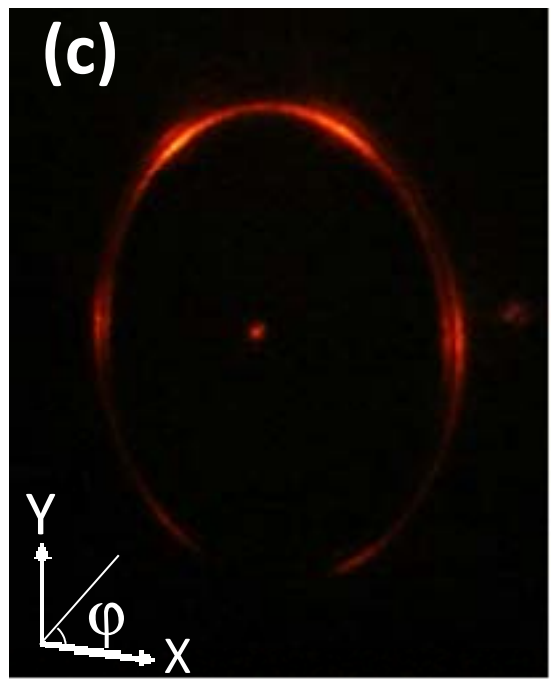
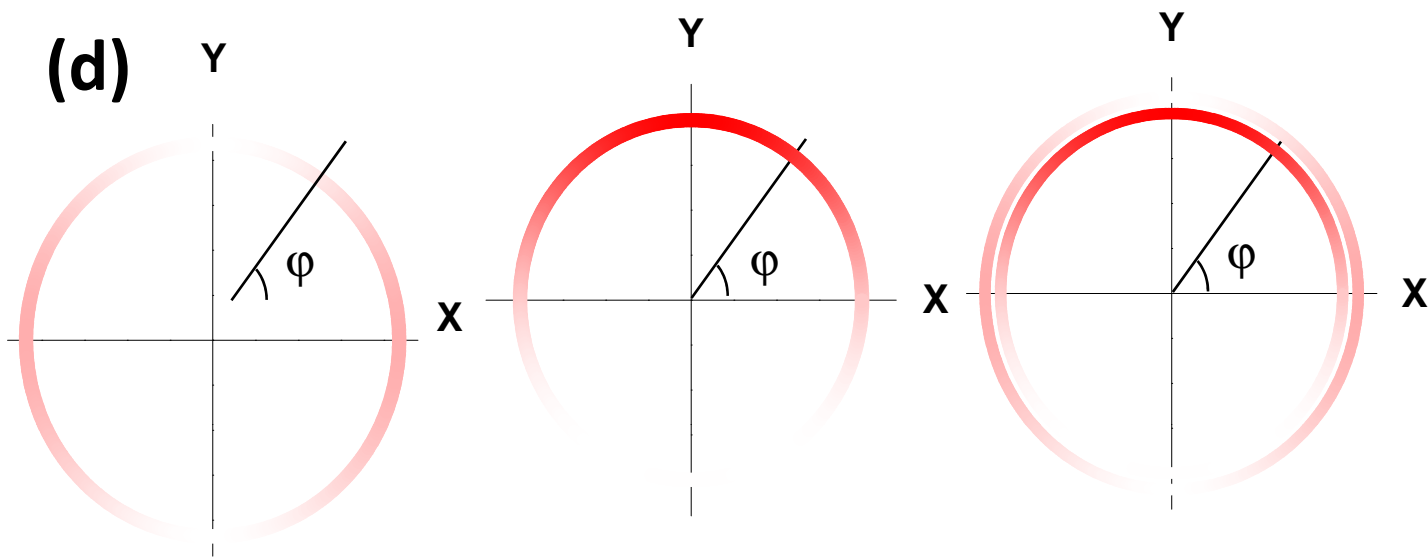

**Figure 1**

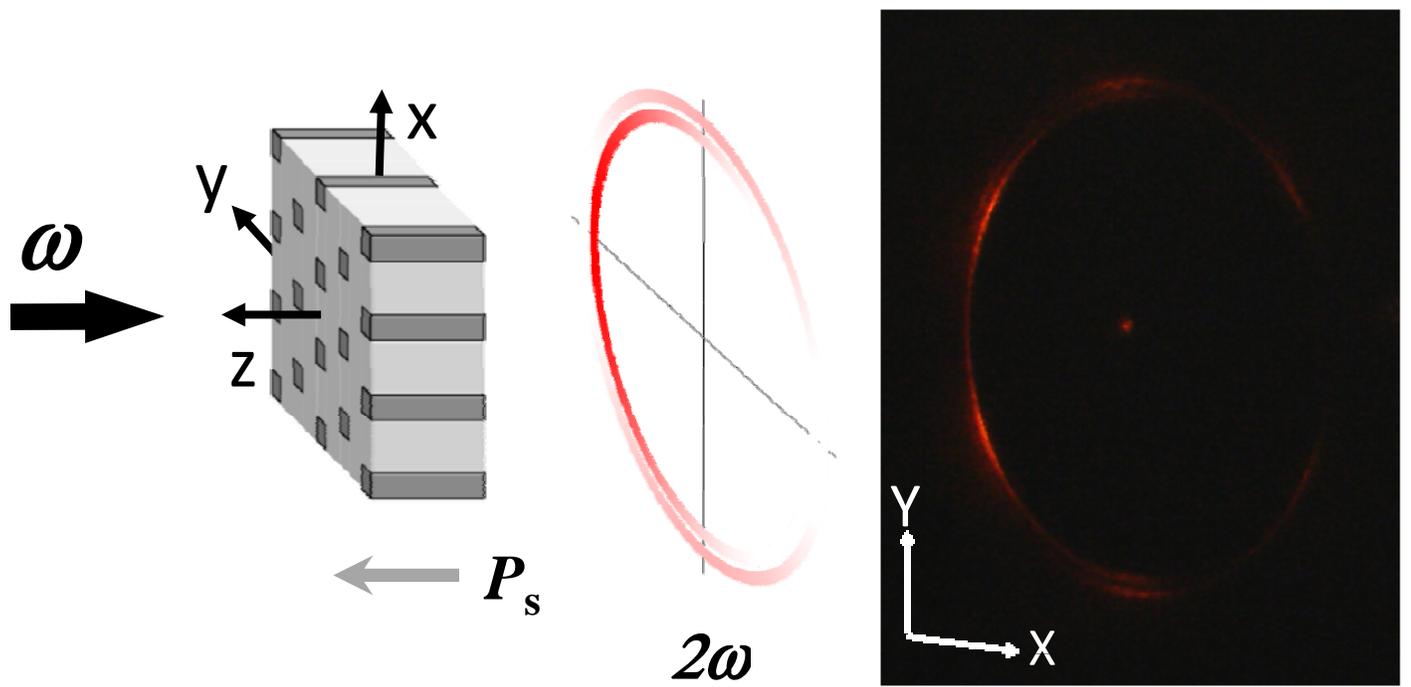

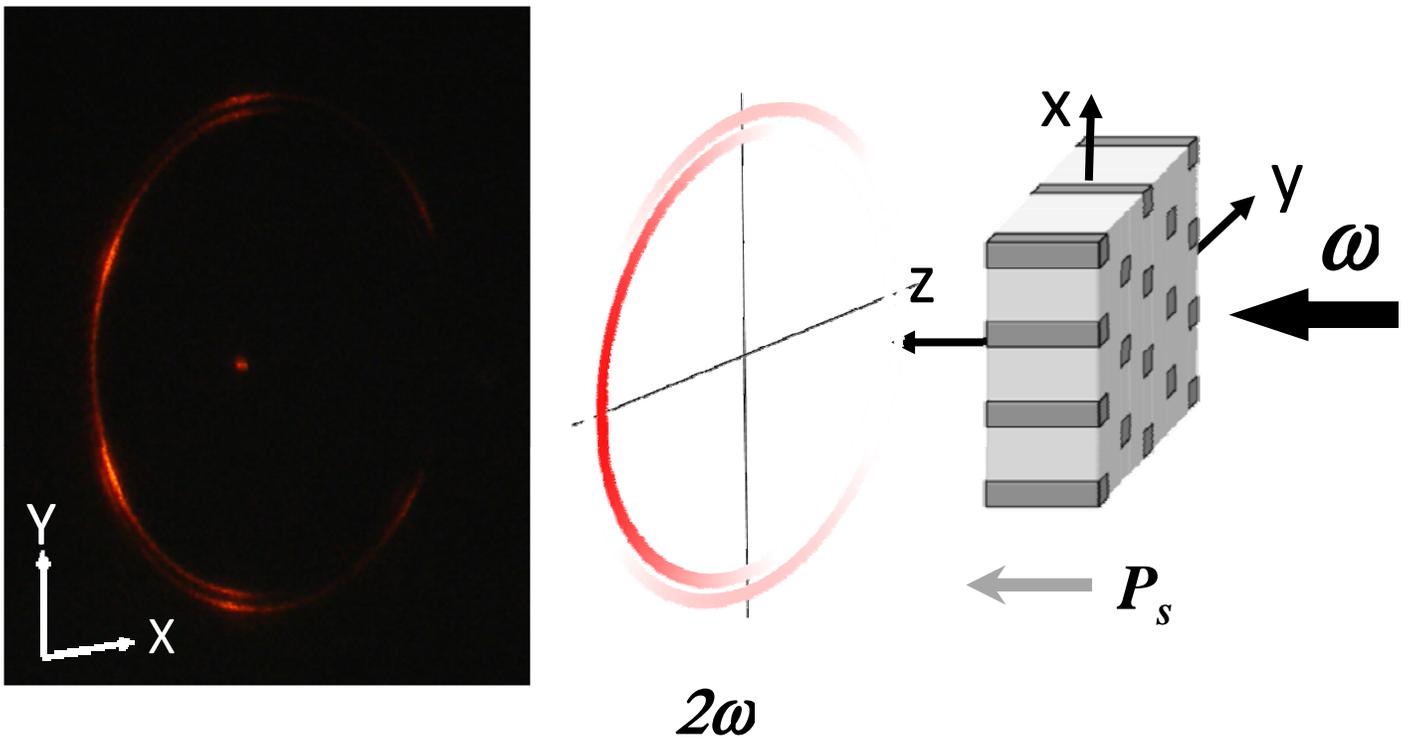

**Figure 2**

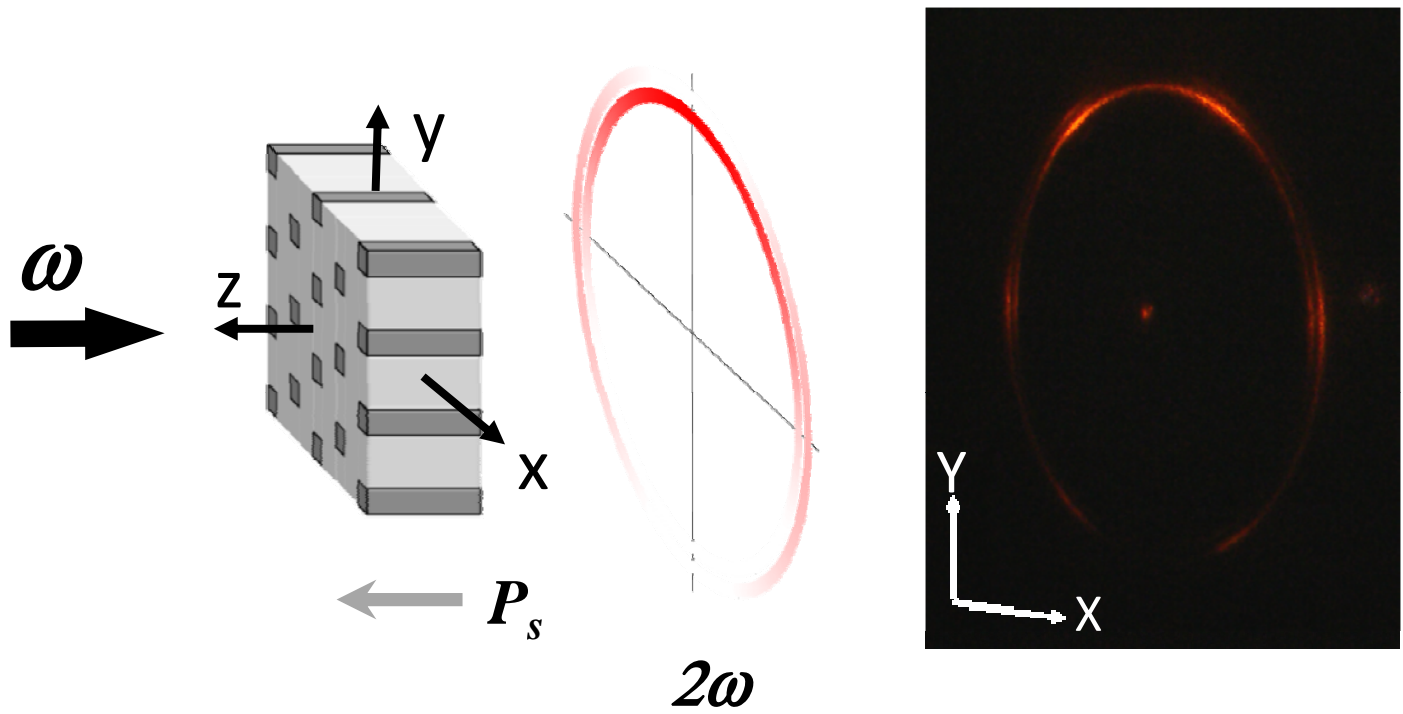
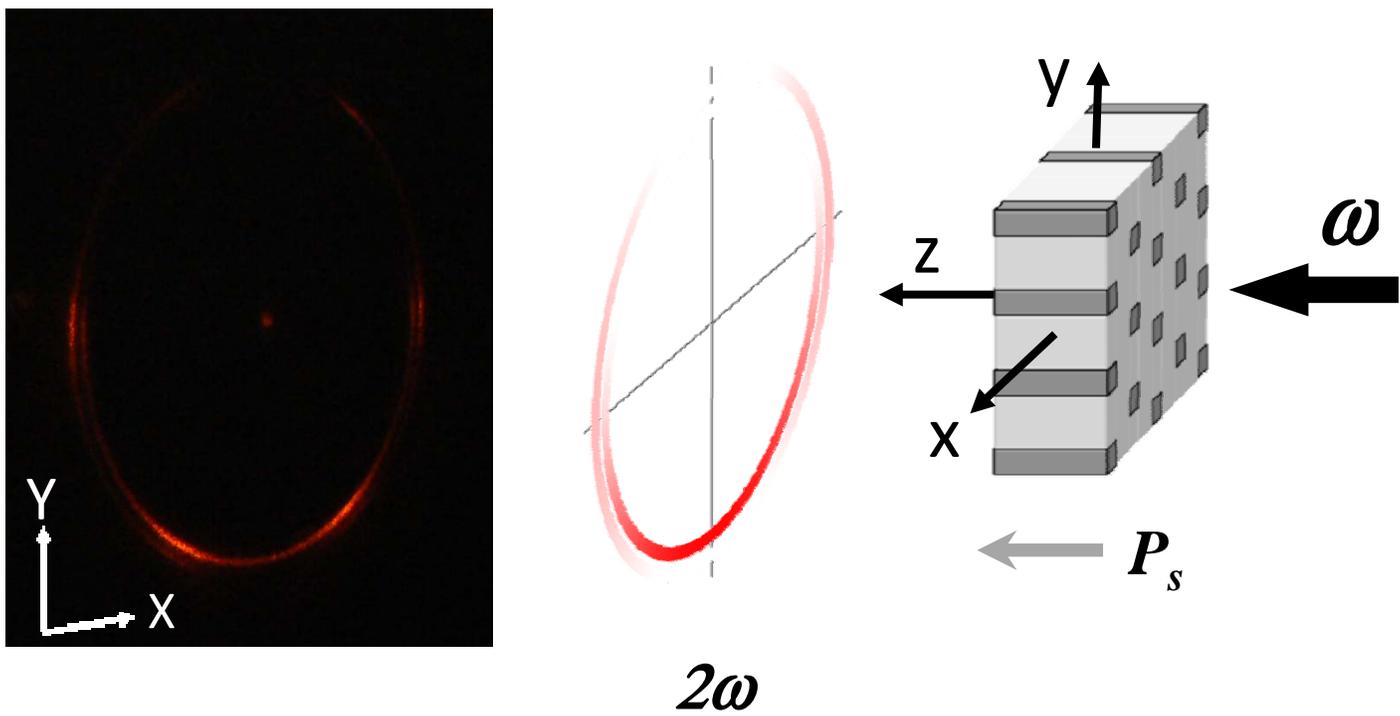

**Figure 3**

**(a)**

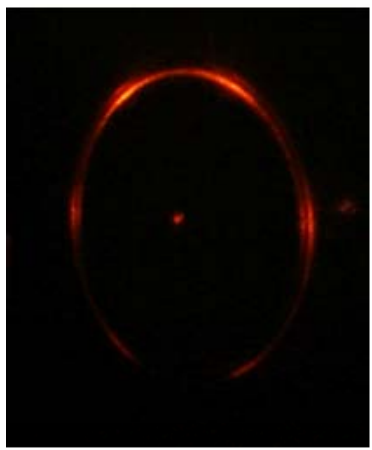 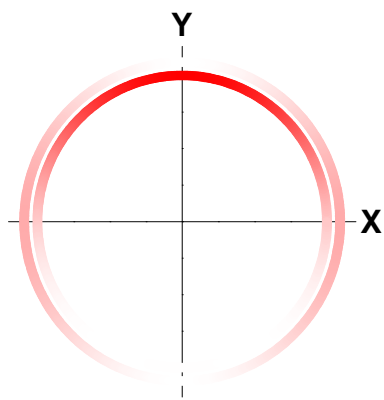

Pol // x
$\gamma = 0°$

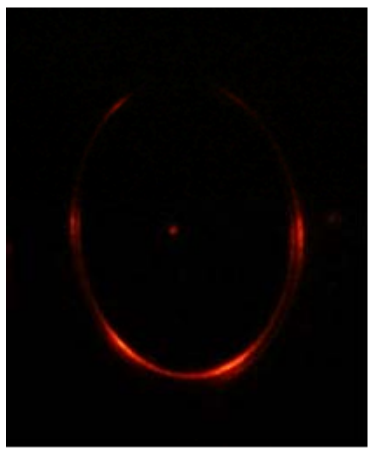 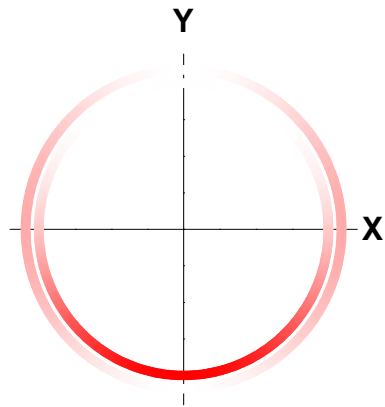

Pol // y
$\gamma = \pi/2$

**(b)**

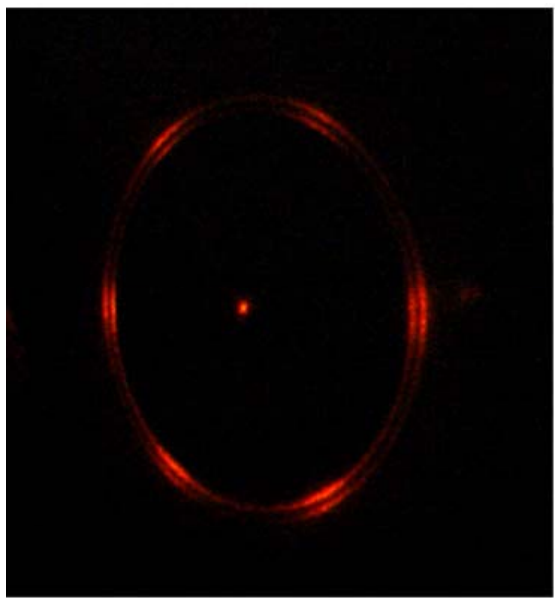 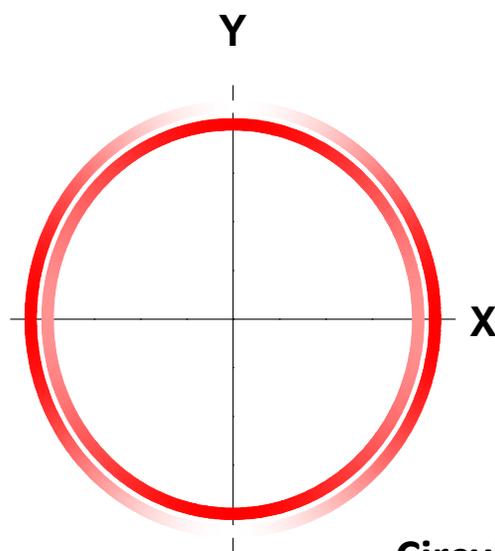

Circular Polarization

**Figure 4**